\documentclass[12pt]{article}

\usepackage{amsmath}
\usepackage{amscd}
\usepackage{amssymb}
\usepackage{amsfonts}
\usepackage[mathscr]{euscript}
\usepackage{pstricks}
\usepackage{pst-node}
\usepackage{graphicx}

\title{\bf Entanglement and entropy rates in open quantum systems}

\author{Fabio Benatti$^{a,b}$,
Alexandra M. Liguori$^{a,b}$, Giacomo Paluzzano\\
\small $^a$Dipartimento di Fisica Teorica, Universit\`a di Trieste,
Strada Costiera 11,\\
\small 34014 Trieste, Italy\\
\small $^b$Istituto Nazionale di Fisica Nucleare, Sezione di
Trieste, 34100 Trieste, Italy}

\date{\null}

\begin{document}

\maketitle

\begin{abstract}
We study a recent conjecture about the behavior of the quantum
relative entropy compared to the relative entropy of entanglement in
open bipartite systems. The conjecture states that, under a
dissipative time-evolution, the positive rate of change of the
relative entropy will always be larger than that of the relative
entropy of entanglement. After explicitly solving a two-qubit master
equation of Lindblad-type with separable and entangled stationary
states, we show that the conjecture can be violated for initial
states with an entangled asymptotic state, while it appears to be
confirmed when the asymptotic states are separable.
\end{abstract}

\section{Introduction}

The importance of quantum entanglement as a physical resource for performing
informational tasks which would be classically impossible~\cite{Bruss} has
spurred the study of its dynamical behavior in many different systems.
The time-evolution of most of these is reversible and generated by a Hamiltonian;
however, it is important for quantum entanglement to be used as an efficient
physical resource that its temporal behavior should also be studied
when systems are driven by noisy environments and their dynamics is
irreversible.

In the following, we will consider \textit{open quantum systems}~\cite{AL,spohn,BP}
i.e. systems where the interactions between the subsystem $S$ and the
external environment $E$, though weak, cannot be neglected.
In this case, a standard way of obtaining a manageable dissipative
time-evolution of the density matrix $\varrho_t$ describing the
state of $S$ at time $t$ is to construct it as the solution of a
Liouville-type master equation $\partial_t \varrho_t =
\mathbf{L}[\varrho_t]$, where the generator $\mathbf{L}$ of Lindblad
type~\cite{GFK,Lindblad} takes care of the effects of the environment through
a characteristic matrix of coefficients known as \textit{Kossakowski matrix}.
This can be done by tracing away the
environment degrees of freedom and by
performing a Markovian approximation, i.e.
by studying the evolution on a slow time-scale and neglecting fast
decaying memory effects. Then, the irreversible reduced dynamics of
$S$ is described by one-parameter semigroups of linear maps,
called \textit{quantum dynamical semigroups},
obtained by exponentiation:
$\gamma_t = e^{t\mathbf{L}}$, $t \geq 0$, such that %$\gamma_{t+s} = \gamma_t
%\circ \gamma_s = \gamma_s \circ \gamma_t$ ($s,t \geq 0$) and
$\varrho_t \equiv \gamma_t[\varrho]$. In order to guarantee full
physical consistency, namely that ${\rm id}\otimes \gamma_t$ be
positivity preserving on all states of the compound system $S+S_d$
for any inert ancilla $S_d$, $\gamma_t$
must be completely positive~\cite{AL,spohn,ben-flore}.

The formalism of open quantum systems has been used to describe the
tendency to thermal equilibrium of a small system in weak
interaction with a large heat bath at a certain temperature. The
main tool in this thermodynamical picture is the quantum relative
entropy~\cite{spohn}; it is related to the difference between the
free energy of the irreversibly evolving open quantum system and
that of its equilibrium asymptotic state: this difference
monotonically decreases in time because so does the quantum relative
entropy with respect to completely positive maps~\cite{OP}, as
quantum dynamical semigroups are. Namely, the time derivative of the
quantum relative entropy, called \textit{entropy rate}, has a
definite sign.

The quantum relative entropy has also been used as a possible measure
of the entanglement content of a quantum state: the so-called
\textit{relative entropy of entanglement} provides a pseudo-distance
between a state and the closed convex set of separable states~\cite{plenio-vedral}.

In~\cite{vedral} the natural question was raised whether the entropy
production due to thermodynamical tendency to equilibrium is
somewhat related to the \textit{entanglement rate}, that is to the
speed of variation of the relative entropy of entanglement. A
conjecture was put forward that for systems immersed in an external
bath without a direct source of entanglement due to Hamiltonian
interactions, the absolute value of the entanglement production is
always smaller than the entropy production.

Typically, a system $S$ immersed in a large environment $E$ is
subjected to decoherence; therefore, one expects quantum
entanglement to be generically depleted by a dissipative and noisy
time-evolution. The conjecture mentioned above is motivated by the
fact that, if a quantum open system tends to a separable equilibrium
state, then, in a suitable neighborhood of the latter, the
entanglement production is zero while the entropy production is not.
Indeed, in~\cite{vedral} a concrete example that validates the
conjecture is offered of a 2-qubit system in which only one of them
evolves as a quantum open system. In such a case an initial
maximally entangled state evolves towards a separable steady state
with an entropy production always larger than the speed with which
entanglement is dissipated.

However, in certain specific situations, an environment affecting
both parties of a bipartite system may even build quantum
correlations between the subsystems which compose $S$ (see,
e.g.,~\cite{beige, braun, jacob, ben-flore, palma}). In particular,
in~\cite{ben-flore} this possibility is shown to depend on the
specific form of the generator of the reduced dynamics.
In~\cite{BFP} an inequality was found, involving the entries of such
a matrix which, if fulfilled, is sufficient to ensure that a
specific initial separable pure state of two qubits gets entangled.
Further, in~\cite{liguori-nagy} this inequality was proven to be a
necessary and sufficient condition for environment-induced
entanglement in an initially separable pure state of two qubits.
Even more interestingly, starting from an initially separable state,
the entanglement generated at small times can persist
asymptotically; also, starting from an initially entangled state,
its entanglement content can asymptotically increase.

This work is organized as follows: in Section 2, we consider the
open dynamics of two qubits with a generator that depends on a
parameter which allows to range over all the above mentioned cases,
analytically solving the master equation; then, in Section 3, we
introduce the notions of entropy and entanglement rates and the
conjecture from~\cite{vedral}; finally, in Section 4, numerically
studying the time-behaviour of the entanglement and entropy rates
for various initial states, we show that, whenever there is
asymptotic entanglement the conjecture in~\cite{vedral} is violated,
while it holds if there is no asymptotic entanglement.

\section{The Reduced Dynamics}

Let a bipartite system composed of two qubits be immersed in an external environment in such a way
that, via standard weak-coupling limit techniques~\cite{AL},
one describes their reduced, irreversible dynamics by means of the master equation
\begin{equation}
\label{master-eq}
\partial_t\varrho_t=\mathbf{L}[\varrho_t] = -i\,\frac{\Omega}{2}\Big[\Sigma_3\,,\,\varrho_t\Big]\, +\,
\sum_{i,j=1}^3 A_{ij}\Big(\Sigma_i \varrho_t \Sigma_j -
\frac{1}{2}\{\Sigma_j\Sigma_i, \varrho_t\}\Big)\ ,
\end{equation}
where $\Omega$ is the system frequency, $\Sigma_i := \sigma_i\otimes
\mathbf{I} + \mathbf{I} \otimes \sigma_i$, $\mathbf{I}$ is the
$2\times2$ identity matrix, $\sigma_i$, $i=1,2,3$ are the Pauli
matrices and the matrix
\begin{equation}
\label{Kossmat0}
A=[A_{ij}]=
\begin{pmatrix}

                 1  & i \alpha & 0 \\

                 -i \alpha & 1 & 0 \\

                 0  & 0 & 1
\end{pmatrix}\ ,\qquad \alpha\in\mathbb{R}\ ,\ \alpha^2\leq 1\ ,
\end{equation}
is positive semi-definite. This latter request ensures that the
semigroup generated by~(\ref{master-eq}) consist of completely
positive maps $\gamma_t$ for all $t\geq 0$~\cite{AL}.
\medskip

\noindent
{\bf Remark 1}\quad
By means of the single qubit Pauli matrices $\sigma_i^{(1)}=\sigma_i\otimes\mathbf{I}$ and
$\sigma_i^{(2)}=\mathbf{I}\otimes\sigma_i$ one writes the purely dissipative contribution to the
generator as~\cite{ben-flore}
\begin{equation}
\label{diss-gen}
\mathbf{D}[\rho_t]=\sum_{i,j=1}^3A_{ij}\sum_{a,b=1}^2\Bigl(\sigma^{(a)}_i\,\rho_t\,\sigma^{(b)}_j\,-\,
\frac{1}{2}\left\{\sigma_j^{(b)}\sigma_i^{(a)}\,,\,\rho_t\right\}\Bigr)\
.
\end{equation}
In this way there are six Kraus operators $\sigma^{(a)}_i$, $a=1,2$,
$i=1,2,3$ and the $6\times 6$ Kossakowski matrix reads
\begin{equation}
\label{Kossmat}
K=[K^{(ab)}_{ij}]=\begin{pmatrix}
K^{(11)}&K^{(12)}\cr
K^{(21)}&K^{(22)}
\end{pmatrix}=\begin{pmatrix}A&A\cr
A&A\end{pmatrix}\ .
\end{equation}
From the theory of open quantum
systems~\cite{AL,GFK,spohn,BP} one knows that the
coefficients $K^{(ab)}_{ij}$ in the Kossakowski matrix relative to
the $i$-th Pauli matrix of the $a$-th qubit, respectively the $j$-th
Pauli matrix of the $b$-th qubit, $a,b=1,2$, $i,j=1,2,3$, are
determined by the Fourier transforms of the two-point
time-correlation functions with respect to an environment
equilibrium state $\omega$, $\omega(B^{(a)}_iB^{(b)}_j(t))$, of the
environment operators $B^{(a)}_i$ appearing in the
system-environment interaction  $H_I=\sum_{i=1}^3\Bigl(\sigma^{(1)}_i\otimes
B^{(1)}_i\,+\,\sigma^{(2)}_i\otimes B^{(2)}_i\Bigr)$. The symmetric
form of~(\ref{diss-gen}) thus results when both qubits are linearly
coupled to bath operators such that:
$B^{(1)}_{1,2,3}=B^{(2)}_{1,2,3}=B_{1,2,3}$ and
$\omega(B_{1,2}B_3(t))=0$.
\medskip

\noindent \textbf{Remark 2}\quad Considering two qubits weakly interacting with
a thermal bath modeled as a collection of spinless, massless scalar
fields (see, e.g.,~\cite{ben-flore}) at very high temperature $T=1/\beta$, the
parameter $\alpha$ in the Kossakowski matrix is related to $\beta$, 
i.e. $\alpha = -\beta\Omega$, where $\Omega$ is the system frequency when isolated
from the environment.
Correspondingly, in the case of one qubit immersed in such a thermal
bath at high temperature ($\beta\ll 1$), any initial state
is driven to the thermal asymptotic state
$$
\rho_\infty=\frac{\exp(-\beta\,\Omega\,\sigma_3)}{2\cosh\beta\Omega}\simeq\frac{1}{2}
\Big(1-\beta\Omega\,\sigma_3\Big)\ .
$$
\medskip

The master equation~(\ref{master-eq}) is explicitly integrated in
Appendix A; in the following we will mainly focus upon the
time-evolution of initial states of the form
\begin{equation}
\label{in-state}
\varrho = a\,|1\rangle\langle 1| + d\,|2\rangle\langle 2|
+ b\,|3\rangle\langle 3| + c\,|4\rangle\langle 4|\ ,\quad
a,b,c,d\in\mathbb{R}^+\ ,\ a+b+c+d=1\ ,
\end{equation}
diagonal with respect to the orthonormal vectors
\begin{equation}
\label{ONB}
\vert 1\rangle=\vert00\rangle\ ,\ \vert 2\rangle=\vert11\rangle\ ,\
\vert 3\rangle=\frac{\vert01\rangle+\vert10\rangle}{\sqrt{2}}\ ,\
\vert 4\rangle=\frac{\vert01\rangle-\vert10\rangle}{\sqrt{2}}\ ,
\end{equation}
where $\sigma_3\vert0\rangle=\vert0\rangle$,
$\sigma_3\vert1\rangle=-\vert1\rangle$, and $\vert 00\rangle$,
$\vert 01\rangle$, $\vert10\rangle$, $\vert11\rangle$ form the
so-called standard basis in $\mathbb{C}^2\otimes\mathbb{C}^2$, with
respect to which the states~(\ref{in-state}) are represented by
\begin{equation}
\label{matform}
\varrho=\begin{pmatrix}

                 a  & 0 & 0 & 0  \\

                 0 & \frac{b+c}{2} & \frac{b-c}{2} & 0 \\

                 0  & \frac{b-c}{2} & \frac{b+c}{2} & 0 \\

                 0 & 0 & 0 & d
\end{pmatrix} \ .
\end{equation}
From equations~(\ref{r00})--(\ref{r33}) and~(\ref{A2}) in Appendix
A, it turns out that these initial states evolve at time $t\geq 0$
into states of the same form
\begin{equation}
\label{rho-tilde}
\varrho_t = a_t\,\vert 1\rangle\langle 1\vert\,+\, d_t\,\vert 2\rangle\langle 2\vert\,+\, b_t\,
\vert 3\rangle\langle 3\vert\,+\,c_t\,\vert 4\rangle\langle 4\vert\ ,
\end{equation}
where $c_t=c$ and
\begin{eqnarray}
\nonumber
a_t&=&\frac{(1-\alpha)^2}{3+\alpha^2}\;R\,+\,\sqrt{1-\alpha^2}\,
\frac{(1+\alpha)^2\,a\,-2(1-\alpha)\,d\,+\,
(1+\alpha)^2\,b}{(1+\alpha)(3+\alpha^2)}\;E_-(t)\\
&+&
\frac{2(1+\alpha)\, a\,-(1-\alpha)^2(b+d)}{3+\alpha^2}\;E_+(t)
\label{sa}
\\
\nonumber
d_t&=&\frac{(1+\alpha)^2}{3+\alpha^2}\;R\,-\,\sqrt{1-\alpha^2}\,\frac{2(1+\alpha)\,a\,
-(1-\alpha)^2(b+d)}{(1-\alpha)(3+\alpha^2)}\;\ E_-(t)\\
&-&
\frac{(1+\alpha)^2\,a\,-2(1+\alpha)\,d\,+(1+\alpha)^2\,b}{3+\alpha^2}\;E_+(t)
\label{sb}\\
\nonumber
b_t&=&\frac{(1-\alpha^2)}{3+\alpha^2}\;R\,+\,\sqrt{1-\alpha^2}\,\frac{(1+\alpha)^3\,a\,
+(1-\alpha)^3\,d\,-2(1-\alpha^2)\,b}{(3+\alpha^2)(1-\alpha^2)}\;E_-(t)\\
\label{sc}
&+&
\frac{2(1+\alpha^2)\,b\,-(1-\alpha^2)(a+d)}{3+\alpha^2}\;E_+(t)\ ,
\end{eqnarray}
with $R=a+b+d=1-c$ and
$$
E_+(t)={\rm e}^{-8t}\,\cosh{4t\sqrt{1-\alpha^2}}\ ,\quad
E_-(t)={\rm e}^{-8t}\,\sinh{4t\sqrt{1-\alpha^2}}\ .
$$

Since $\lim_{t\to+\infty}E_\pm(t)=0$, the asymptotic states resulting from the
initial states~(\ref{in-state}) are
\begin{equation}
\label{rhoinfty}
\varrho_\infty(c) =\frac{(1-\alpha)^2}{3+\alpha^2}\,(1-c)\,\vert 1\rangle\langle 1\vert\,+\,
\frac{(1+\alpha)^2}{3+\alpha^2}\,(1-c)\,\vert 2\rangle\langle 2\vert\,+\,
\frac{(1-\alpha^2)}{3+\alpha^2}\,(1-c)\,\vert 3\rangle\langle 3\vert\,+\, c\,\vert 4\rangle\langle 4\vert\ .
\end{equation}
There is thus a one-parameter family $\{\varrho_\infty(c)\}_{0\leq
c\leq 1}$ of asymptotic states such that all initial states of the
form~(\ref{in-state}) with the same $c$ go into the same
$\varrho_\infty(c)$.

In order to study the asymptotic entanglement generation capability
of the present model, we shall measure the entanglement of 2-qubit
states $\rho$ by the concurrence~\cite{Wootters}:
$$
C(\varrho)=\max\{0,\lambda_1-\lambda_2-\lambda_3-\lambda_4\}\ ,
$$
where $\lambda_1\geq\lambda_2\geq\lambda_3\geq\lambda_4\geq 0$ are
the square roots of the positive eigenvalues of
$\varrho\widetilde{\varrho}$ with
$\widetilde{\varrho}=\sigma_2\otimes\sigma_2\,\varrho^*\,\sigma_2\otimes\sigma_2$,
$\varrho^*$ denoting the complex conjugated matrix. For 2-qubit
states of the form~(\ref{matform}), $\rho=\rho^*$ and one easily computes
$$
\widetilde{\varrho}=
\begin{pmatrix}
d&0&0&0\cr
0&\frac{b+c}{2}&\frac{b-c}{2}&0\cr
0&\frac{b-c}{2}&\frac{b+c}{2}&0\cr
0&0&0&a
\end{pmatrix}\ ,\quad
\varrho\widetilde{\varrho}=
\begin{pmatrix}
ad&0&0&0\cr
0&\frac{b^2+c^2}{2}&\frac{b^2-c^2}{2}&0\cr
0&\frac{b^2-c^2}{2}&\frac{b^2+c^2}{2}&0\cr
0&0&0&ad
\end{pmatrix}\ .
$$
This latter matrix has positive eigenvalues $ad$ (twice degenerate), $b^2$, $c^2$;
then, their square roots $\lambda_1\geq\lambda_2\geq\lambda_3\geq\lambda_4$
in decreasing order yield
\begin{eqnarray}
\label{concurrence1}
C(\varrho)&=&\max\left\{0,2\left(\frac{|b-c|}{2}-
\sqrt{ad}\right)\right\}\\
C(\varrho_\infty)&=&\max\left\{0,\frac{\Big|1-\alpha^2-4c\Big|-2(1-\alpha^2)(1-c)}{3+\alpha^2}\right\}
\ .
\end{eqnarray}

\noindent \textbf{Remark 3}\quad As already emphasized in the
introduction, despite decoherence, the presence of an environment
need not have only destructive effects in relation to entanglement:
entanglement can even be asymptotically increased with respect to
the initial amount. This can happen in the present case and the
entanglement generation capability of the environment is entirely
due to the non-Hamiltonian contribution~(\ref{diss-gen}) to the
generator in~(\ref{master-eq}). Indeed, the 2-qubit Hamiltonian does
not contain coupling terms and cannot be a source of entanglement;
instead, this can be true for~(\ref{diss-gen}) because the
off-diagonal contributions in the Kossakowski matrix~(\ref{Kossmat})
couple the two qubits. Of course, this is only necessary, but not
sufficient to ensure entanglement generation and its asymptotic
persistence. They indeed depend on a trade-off between the
off-diagonal couplings  and the purely decohering diagonal terms
in~(\ref{Kossmat}).

\section{Entropy and Entanglement Rates}

In this section we shall introduce the notions of entropy and
entanglement rates; for sake of simplicity, we shall consider finite
$d$-level systems whose states are described by normalized,
positive, $d\times d$ density matrices $\varrho\in M_d(\mathbb{C})$.
Given two such density matrices, their quantum relative entropy is
defined by~\cite{OP}
\begin{equation}
\label{relent}
S(\rho_1||\rho_2)={\rm Tr}\Bigl(\varrho_1(\log\varrho_1-\log\varrho_2)\Bigr)\ .
\end{equation}
Consider a quantum system with Hamiltonian $H$; if in contact
with a heat bath at temperature $T=1/\beta$ (with the Boltzmann constant $\kappa=1$), it is expected to be driven
asymptotically into the thermal (Gibbs) state $\varrho_T =
e^{-\beta H}/Z_\beta$, where $Z_\beta={\rm Tr}[e^{-\beta H}]$.
Suppose that under an irreversible time-evolution $\varrho\mapsto\varrho_t$, an initial state
$\varrho$ is driven into thermal equilibrium, that is
$\lim_{t\to+\infty}\varrho_t=\varrho_T$; then,
$$
\frac{1}{\beta}\,
S(\varrho_t||\varrho_T)=\frac{1}{\beta}{\rm Tr}\Bigl(\varrho_t(\log\varrho_t+\log Z_\beta+\beta\,H)\Bigr)=-T\,S(\varrho_t)\,+\,{\rm Tr}(\varrho_t\,H)\,+T\,\,\log Z_\beta\ ,
$$
where $S(\varrho)=-{\rm Tr}\varrho\log\varrho$ is the von Neumann
entropy of the state $\rho$. Since the second term corresponds to
the system's internal energy, the first two contributions give the
system's free energy
corresponding to the time-evolving state $\rho_t$~\cite{spohn}:
$$
F(\varrho_t)=U(\varrho_t)-T\,S(\varrho_t)\ ,\quad U(\varrho_t)={\rm Tr}(\varrho_t\, H)\ .
$$
Finally, $F(\varrho_T)=-\log Z_\beta$ implies that the quantum relative entropy is related to the difference
of free energies
$$
\label{freenergy1}
S(\varrho_t||\varrho_T)=\beta\Big(F(\varrho_t)\,-\,F(\varrho_T)\Big)\ .
$$
Because of the second law of thermodynamics, the above quantity
should be positive and its time-derivative non-positive. The first
property is guaranteed by the properties of the quantum relative
entropy~\cite{OP}, while the second one holds true when the
irreversible time-evolution is given by a Markovian
semigroup, that is when $\varrho_t=\gamma_t[\varrho]$ and $\gamma_t\circ\gamma_s=\gamma_s\circ\gamma_t=\gamma_{s+t}$ for all $s,t\geq 0$. Indeed, since
$\gamma_t[\varrho_T]=\varrho_T$, one derives
\begin{eqnarray*}
S(\varrho_t\,||\,\varrho_T)&=&S\Big(\gamma_t[\varrho]\,||\,\gamma_t[\varrho_T]\Big)=
S\Big(\gamma_{t-s}\circ\gamma_s[\varrho]\,||\,\gamma_{t-s}\circ\gamma_s[\varrho_T]\Big)\\
&\leq&
S(\gamma_s[\varrho]\,||\,\gamma_s[\varrho_T])=S(\varrho_s\,||\,\varrho_T)\quad\forall\ 0\leq s\leq t\ ,
\end{eqnarray*}
where the last inequality follows from the fact that the quantum
relative entropy decreases under the action of completely positive
trace-preserving maps~\cite{OP}.

Based on the previous thermodynamical arguments, one may consider
generic open quantum dynamics
$\varrho\mapsto\gamma_t[\varrho]=\varrho_t$ with asymptotic states
$\lim_{t\to+\infty}\varrho_t=\varrho_\infty$, that are not
necessarily thermal ones. The speed of convergence to such
stationary states starting from an initial state $\varrho$ will then
be measured by the \textit{entropy rate}
\begin{equation}
\label{entroprod0}
\sigma[\varrho_t] = -\frac{d}{dt}S(\varrho_t||\varrho_\infty)
={\rm Tr}\Big(\dot{\varrho}_t(\log\varrho_\infty\,-\,\log\varrho_t)\Big)\ .
\end{equation}

The entropy production that accompanies the tendency to equilibrium of the states of the form~(\ref{rho-tilde})
is easily computed; indeed,
being the states $\varrho_t$ and $\varrho_\infty$ diagonal with respect to the same orthonormal basis,
the entropy rate~(\ref{entroprod0}) has the analytic expression
\begin{equation}
\label{entroprod1}
\sigma[\varrho_t]=\dot{a}_t\,\log\frac{(1-\alpha)^2(1-c)}{a_t(3+\alpha^2)}\,+\,
\dot{b}_t\,\log\frac{(1-\alpha^2)(1-c)}{b_t(3+\alpha^2)}\,
+\,\dot{d}_t\,\log\frac{(1+\alpha)^2(1-c)}{d_t(3+\alpha^2)}\ .
\end{equation}

When the density matrix $\rho$ is the state of, say, a bipartite quantum system,
it makes sense to introduce the \textit{relative entropy of entanglement},
\begin{equation}
\label{ent-entr0}
E[\varrho]= \inf_{\varrho_{sep}} S(\varrho||\varrho_{sep})\ ,
\end{equation}
as a measure of the entanglement content of $\rho$. Indeed, the
above quantity vanishes if and only if $\rho$ is separable and can
be used to measure the distance of $\rho$~\footnote{The relative entropy of entanglement is not exactly a distance since it is not symmetric.} from
the convex set of separable states; furthermore, it cannot increase,
but at most remain constant, under the action of local operations,
described by trace-preserving completely positive maps acting
independently on the two parties~\cite{vprk, vedral-plenio}.

Analogously to what was done for the entropy production, one may
look at the \textit{entanglement rate} when the system evolves, i.e.
at the time-derivative of the pseudo-distance
\begin{equation}
\label{ent-prod}
\sigma_E[\varrho_t] = \frac{d}{dt}E[\varrho_t] \ .
\end{equation}
In~\cite{vedral} it was argued that
\begin{equation}
\label{vedralconj}
\Big|\sigma_E[\varrho_t]\Big| \leq \sigma[\varrho_t]\
\end{equation}
always holds in absence of direct entangling interactions between
the parties. The argument on which the conjecture is based is that
decoherence is expected to deplete entanglement before reaching the
asymptotic state and thus before the entropy production vanishes.
Such asymptotic intuition is then extrapolated at all times.

Now we will illustrate the various possibilities offered by the reduced dynamics discussed in the previous section; in particular, we will compare
the entropy and entanglement rates, thus checking the validity of the
conjecture~(\ref{vedralconj}).

We will first derive an explicit expression for the relative entropy of
entanglement~(\ref{ent-entr0}) in the case of states of the form~(\ref{rho-tilde})
and then compute numerically the behaviour of its time-derivative~(\ref{ent-prod}).

Let us first rewrite~(\ref{ent-entr0}) as follows:
\begin{equation}
\label{constr0}
E(\varrho_t)=-S(\varrho_t) - \sup_{\varrho_{sep}}{\rm Tr}\Big(\varrho_t\log\varrho_{sep}\Big)\ ,
\end{equation}
where $S(\varrho_t)$ is the von Neumann entropy of the time-evolving state.
The following result, which we prove in Appendix B, helps to explicitly solve the above maximization problem.
\medskip

\noindent
\textbf{Proposition}\quad
In the case of states as in~(\ref{rho-tilde}), the supremum in~(\ref{constr0}) is achieved
for separable states of the form
\begin{equation}
\label{constr1}
\varrho_{sep} = x |1\rangle\langle1| + u
|3\rangle\langle3| + v |4\rangle\langle4| + y |2\rangle\langle2|\ ,
\end{equation}
where the parameters $x,y,u,v$ are real and such that
\begin{equation}
\label{param}
x+u+v+y=1\ ,\quad \quad \frac{|u-v|}{2} \leq \sqrt{x
y}\ .
\end{equation}
This leads to the following maximization problem
\begin{equation}
\label{max}
E(\varrho_t)=-S(\varrho_t)-\sup_{\varrho\in\mathcal{S}_{sep}^{diag}}\Big(
a_t\log x + d_t \log y + b_t \log u + c_t \log v\Big)\ ,
\end{equation}
which can be analytically solved.

\section{Results}

The above maximization problem is explicitly solved in Appendix B thus
permitting to calculate numerically the entanglement
rate~(\ref{ent-prod}), to compare it with the entropy
rate~(\ref{entroprod0}) and to check the
conjecture~(\ref{vedralconj}). We shall do this in a number of cases
that cover all possible initial and asymptotic entanglement
properties for which we plot the behaviors of the relative entropy
and of the relative entropy of entanglement, separately, while the
entropy rate and the entanglement rate are plotted together for
direct comparison.

In the following, the choice of the range of values for the plots'
axes was made only for graphic reasons to make the plots clearer.
Moreover, in the first four cases,  we took the parameter in the
matrix $A$ to be $\alpha=0.5$ as it makes the plots easier to read;
changing $\alpha$ does not alter the results. In the last example,
instead, we show two different behaviors of the entanglement of the
initial state depending on the choice of the parameter $\alpha$.

\noindent \textbf{Case 1.}\quad An initial pure separable
state~(\ref{state1}) goes into a mixed separable state; the
dissipative time-evolution is not able to generate entanglement at
any time, as shown by the second and the third plot below where the entropy of entanglement and the entanglement rate are both zero. In this case the conjecture~(\ref{vedralconj}) holds.
\begin{eqnarray}
\label{state1}
&&
\varrho=|1\rangle\langle 1|\\
\nonumber
&&
\varrho_\infty=
\frac{(1-\alpha)^2}{3+\alpha^2}\,\vert 1\rangle\langle 1\vert\,+\,
\frac{(1+\alpha)^2}{3+\alpha^2}\,\vert 2\rangle\langle 2\vert\,+\,
\frac{(1-\alpha^2)}{3+\alpha^2}\,\vert 3\rangle\langle 3\vert\\
\nonumber
&&
C(\varrho)=C(\varrho_\infty)=0
\end{eqnarray}

\begin{figure}[h!]
\centering
\includegraphics[width=6cm]{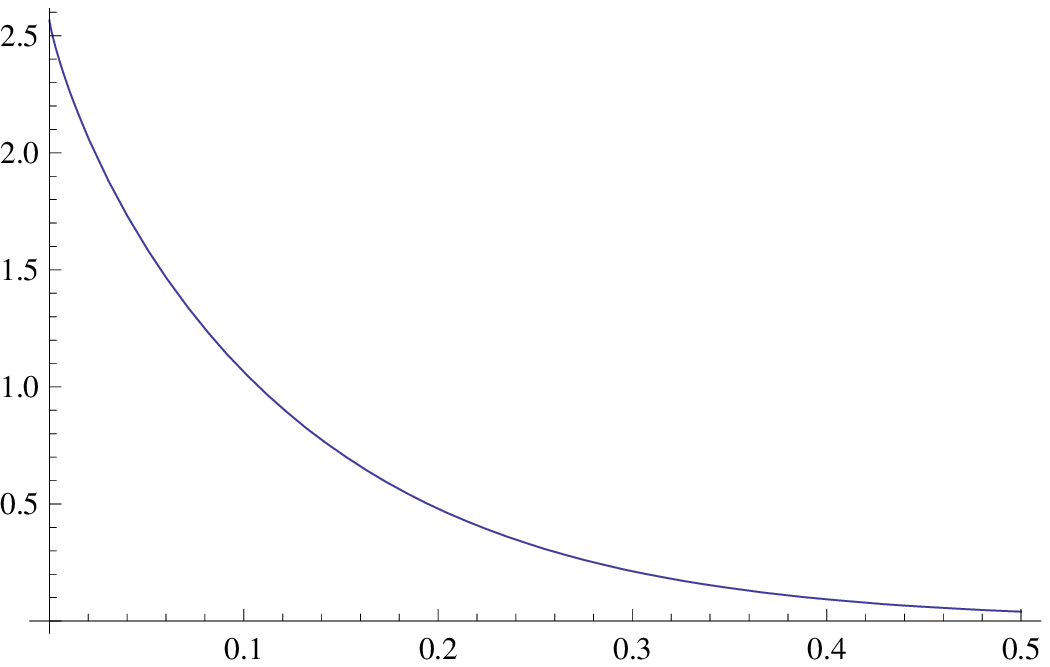}
\includegraphics[width=6cm]{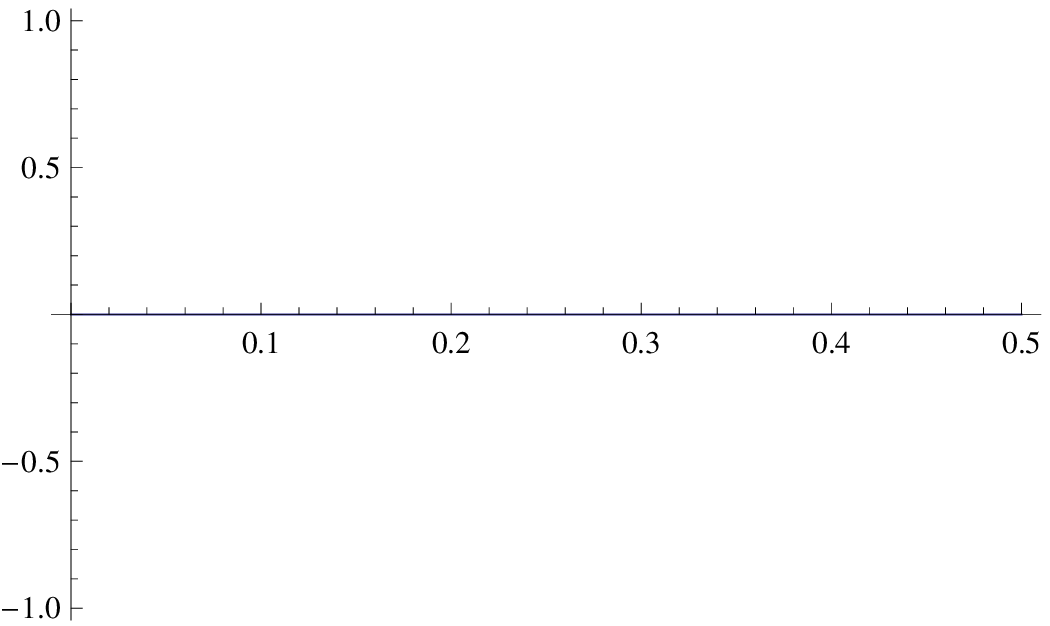}
\caption{Case 1: $\alpha=0.5$; Left: $S(\varrho_t)$; Right:
$E(\varrho_t)$} \label{fig1-1}
\end{figure}

\begin{figure}[h!]
\centering
\includegraphics[width=6cm]{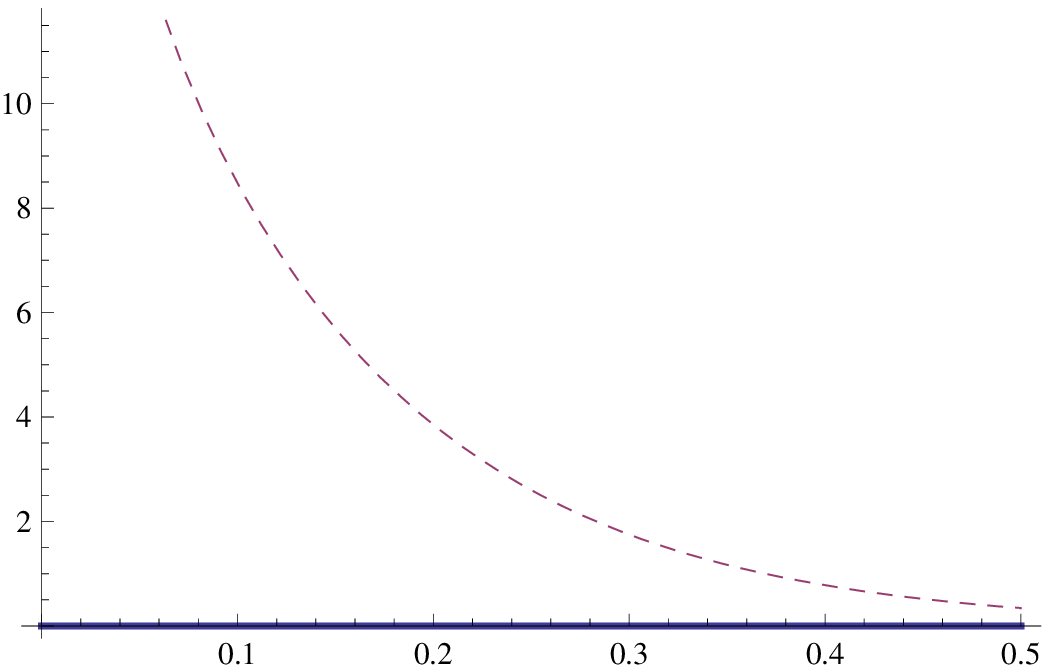}
\caption{Case 1: $\alpha=0.5$; $\sigma[\varrho_t]$ dashed line,
$|\sigma_E[\varrho_t]|$ continuous line} \label{fig1}
\end{figure}

\newpage
\noindent \textbf{Case 2.}\quad An initial mixed separable
state~(\ref{state2}) goes into a mixed entangled state and the
conjecture~(\ref{vedralconj}) is violated after some time.
\begin{eqnarray}
\label{state2}
&&
\varrho=
\frac{1}{2}\,|3\rangle\langle 3|\, +\, \frac{1}{2}\,|4\rangle\langle 4|\\
\nonumber
&&\varrho_\infty=
\frac{(1-\alpha)^2}{2(3+\alpha^2)}\,\vert 1\rangle\langle 1\vert\,+\,
\frac{(1+\alpha)^2}{2(3+\alpha^2)}\,\vert 2\rangle\langle 2\vert\,+\,
\frac{(1-\alpha^2)}{2(3+\alpha^2)}\,\vert 3\rangle\langle 3\vert\,+\,\frac{1}{2}\,\vert 4\rangle\langle 4\vert\\
\nonumber
&&
C(\varrho)=0\ ,\quad C(\varrho_\infty)=\frac{2\alpha^2}{3+\alpha^2}\geq 0 \qquad\forall\alpha\in[-1,1]\ .
\end{eqnarray}

\begin{figure}[h!]
\centering
\includegraphics[width=6cm]{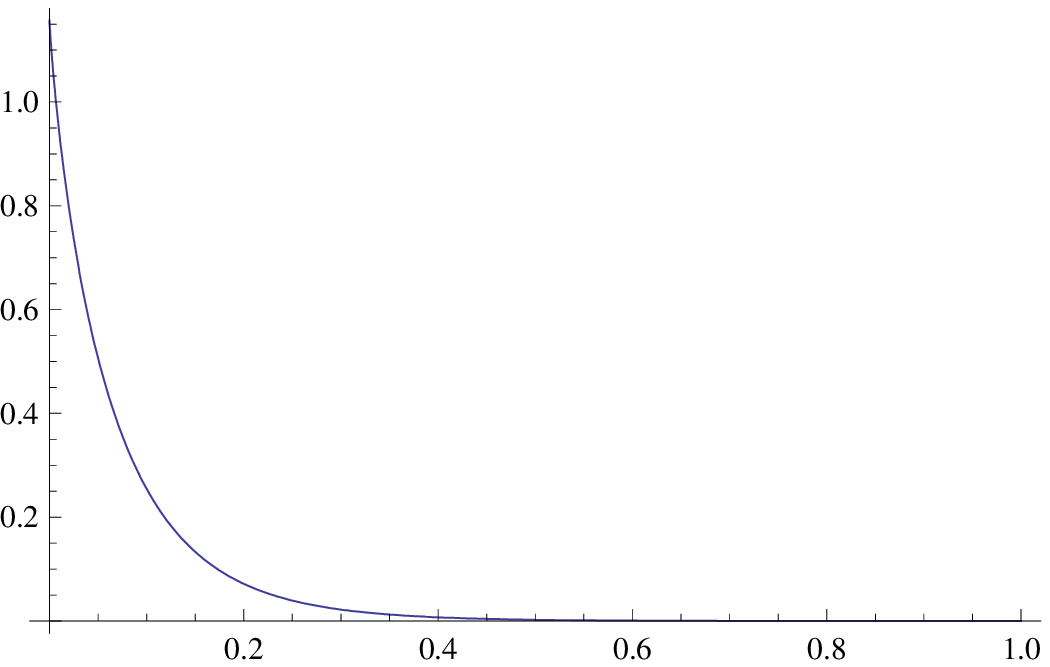}
\includegraphics[width=6cm]{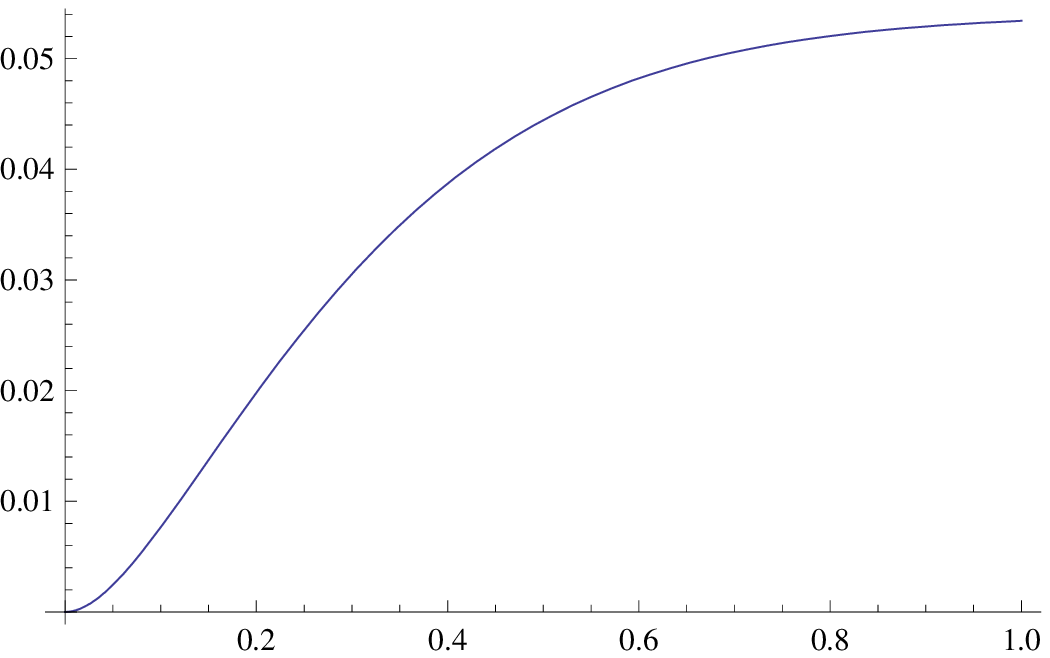}
\caption{Case 2: $\alpha=0.5$; Left: $S(\varrho_t||\varrho_\infty)$; Right:
$E[\varrho_t]$} \label{fig25}
\end{figure}

\begin{figure}[h!]
\centering
\includegraphics[width=6cm]{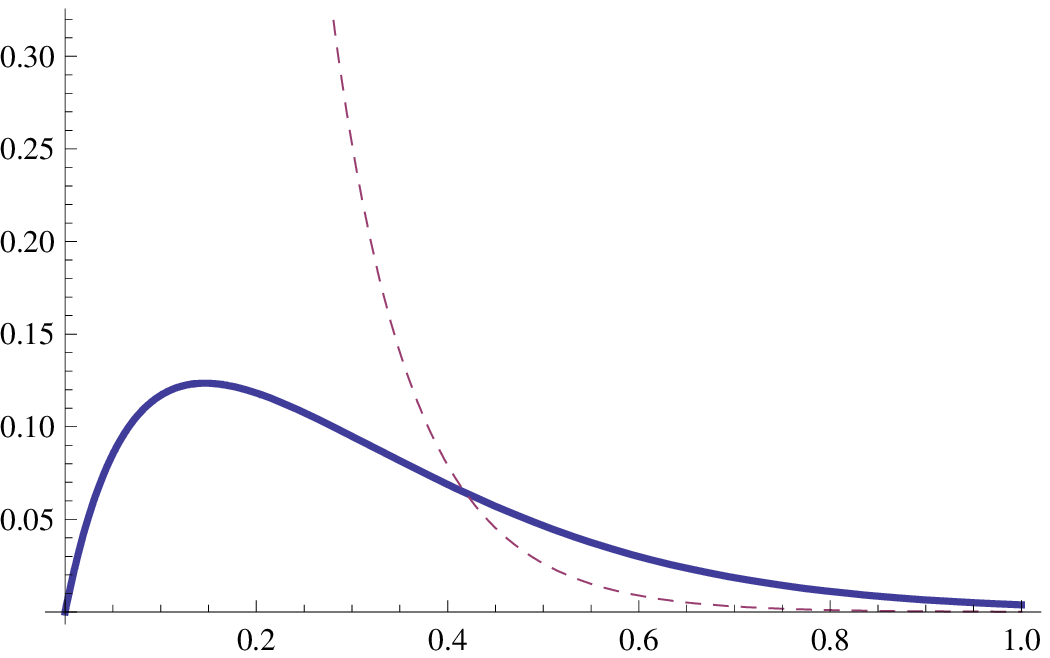}
\caption{Case 2: $\alpha=0.5$; $\sigma[\varrho_t]$ dashed line,
$|\sigma_E[\varrho_t]|$ continuous line} \label{fig2}
\end{figure}

\noindent \textbf{Case 3.}\quad An initial mixed entangled
state~(\ref{state3}) goes into an asymptotic mixed state which is
more or equally entangled and the conjecture~(\ref{vedralconj}) is
violated after some time.
\begin{eqnarray}
\label{state3}
&&
\varrho=
\frac{1}{10}\,\vert 1\rangle\langle 1\vert\,+\,\frac{1}{10}\,\vert 2\rangle\langle 2\vert\,+\,
\frac{1}{10}\,\vert 3\rangle\langle 3\vert\,+\,\frac{7}{10}\,\vert 4\rangle\langle 4\vert\\
\nonumber
&&
\varrho_\infty=
\frac{3(1-\alpha)^2}{10(3+\alpha^2)}\,\vert 1\rangle\langle 1\vert\,+\,
\frac{3(1+\alpha)^2}{10(3+\alpha^2)}\,\vert 2\rangle\langle 2\vert\,+\,
\frac{3(1-\alpha^2)}{10(3+\alpha^2)}\,\vert 3\rangle\langle 3\vert\,+\,\frac{7}{10}\,\vert 4\rangle\langle 4\vert\\
\nonumber && C(\varrho)=\frac{2}{5}\ ,\quad
C(\varrho_\infty)=\frac{2}{5}\frac{3+4\alpha^2}{3+\alpha^2}\geq\frac{2}{5}\qquad\forall\alpha\in[-1,1]\ .
\end{eqnarray}

\begin{figure}[h!]
\centering
\includegraphics[width=6cm]{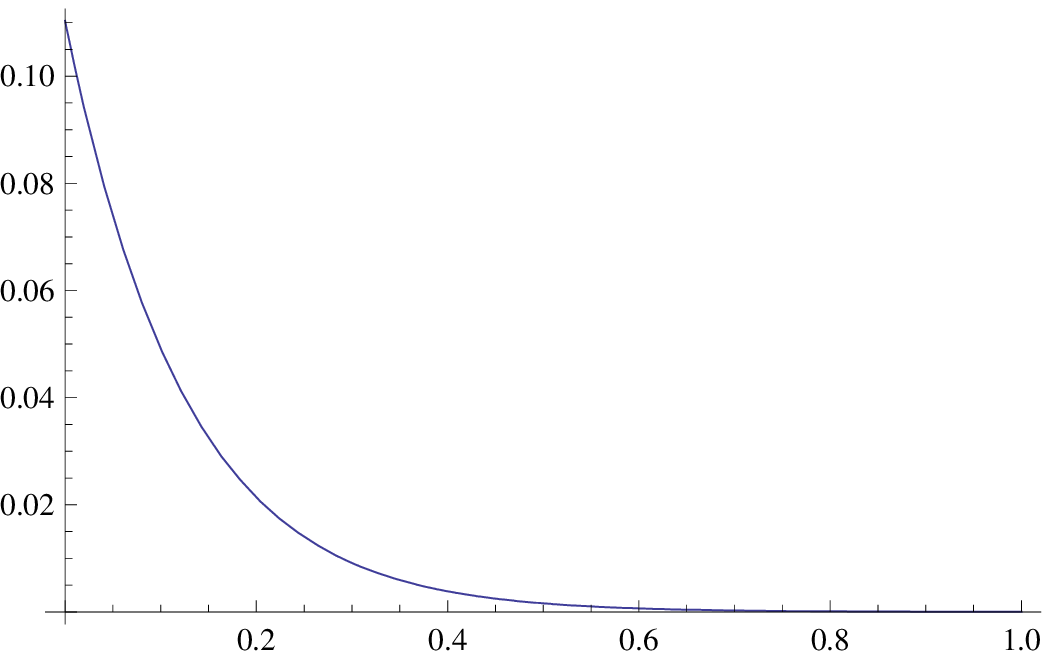}
\includegraphics[width=6cm]{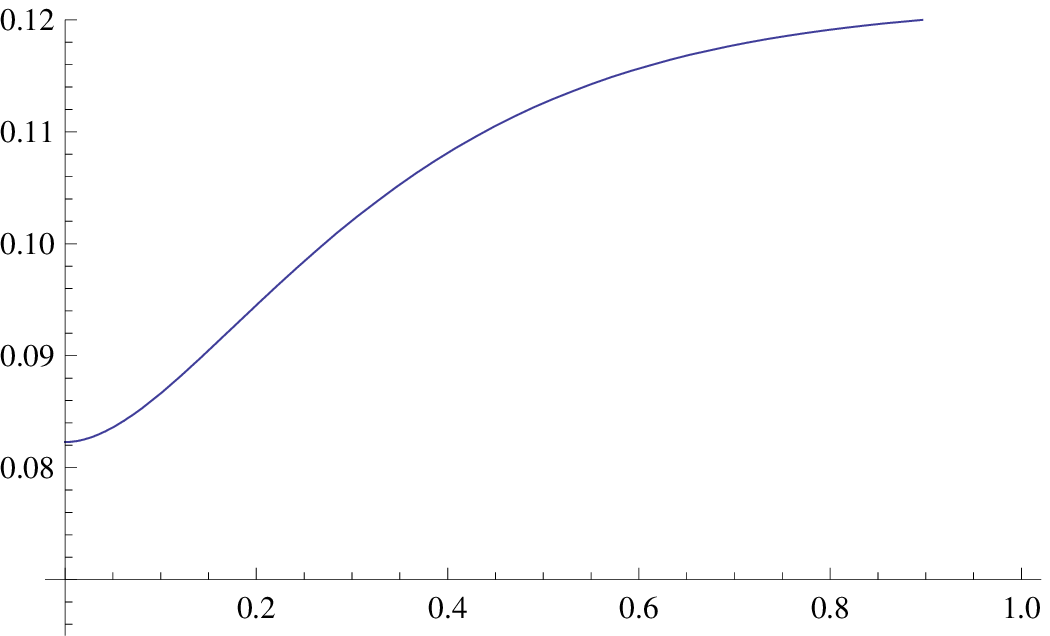}
\caption{Case 3: $\alpha=0.5$; Left: $S(\varrho_t||\varrho_\infty)$; Right:
$E[\varrho_t]$} \label{fig22}
\end{figure}
\begin{figure}[h!]
\centering
\includegraphics[width=6cm]{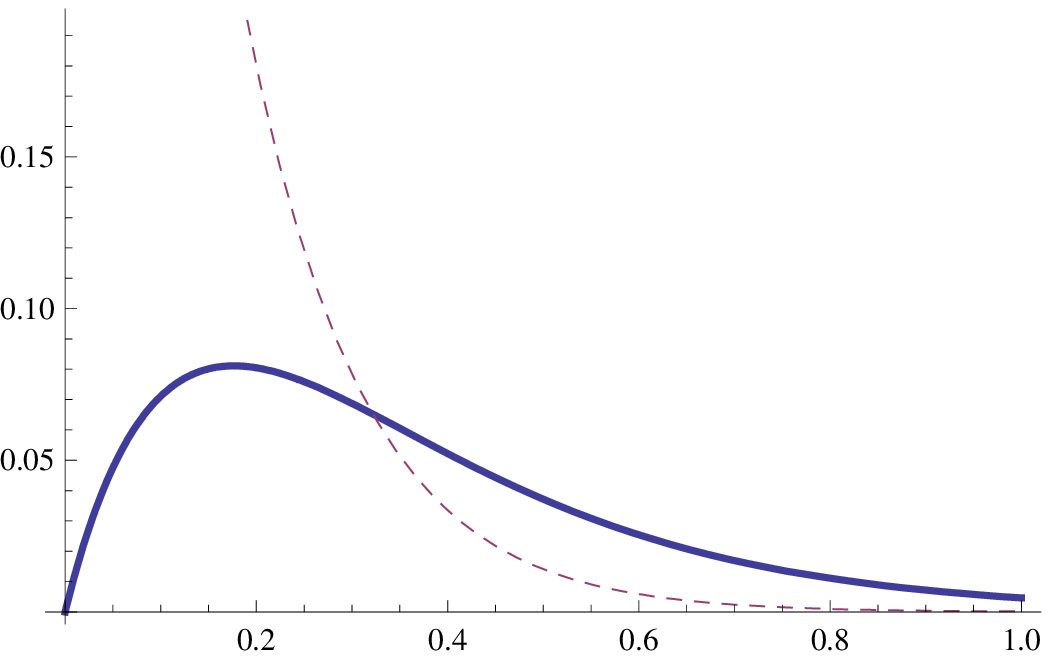}
\caption{Case 3: $\alpha=0.5$; $\sigma[\varrho_t]$ dashed line,
$|\sigma_E[\varrho_t]|$ continuous line} \label{fig3}
\end{figure}

\newpage
\noindent \textbf{Case 4.}\quad An initial mixed entangled
state~(\ref{state5}) goes into a state with less entanglement
\begin{eqnarray}
\label{state5} && \varrho=\frac{1}{2}\vert 2\rangle\langle
2\vert\,+\,\frac{1}{10}\,\vert 3\rangle\langle
3\vert\,+\,\frac{2}{5}\,
\vert 4\rangle\langle 4\vert\\
\nonumber && \varrho_\infty=
\frac{3(1-\alpha)^2}{5(3+\alpha^2)}\,\vert 1\rangle\langle
1\vert\,+\, \frac{3(1+\alpha)^2}{5(3+\alpha^2)}\,\vert
2\rangle\langle 2\vert\,+\,
\frac{3(1-\alpha^2)}{5(3+\alpha^2)}\,\vert 3\rangle\langle 3\vert\,+\,\frac{2}{5}\,\vert 4\rangle\langle 4\vert\\
\nonumber && C(\varrho)=\frac{3}{5}\ ,\quad
\left\{\begin{matrix}C(\varrho_\infty)=0 \quad \text{for}\;\alpha^2
\leq\frac{3}{11}\cr
C(\varrho_\infty)=\frac{11\alpha^2-3}{5(3+\alpha^2)}<\frac{3}{5}
\quad \text{for}\;\frac{3}{11}<\alpha^2\leq 1
\end{matrix}\right.\ .
\end{eqnarray}

With the choice $\alpha=0.5$, the dissipative time-evolution shows a
sudden death of entanglement, that is the concurrence\footnote{In Figure $7$,
instead of plotting the concurrence as defined
in~(\ref{concurrence1}), we have plotted the difference
$|b-c|-2\sqrt{ad}$: this simply means that, as soon as this
difference becomes negative, the state is
separable.}~(\ref{concurrence1}) vanishes at finite time.
The conjecture~(\ref{vedralconj}) always holds.

\begin{figure}[h!]
\centering
\includegraphics[width=6cm]{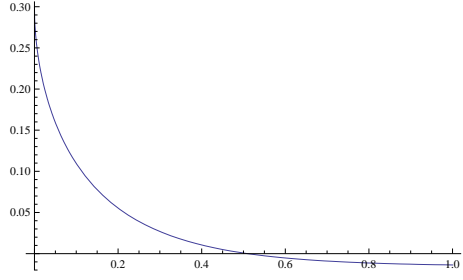}
\caption{Case 4: $\alpha=0.5$; $C(\varrho_t)$} \label{fig5}
\end{figure}

\begin{figure}[h!]
\centering
\includegraphics[width=6cm]{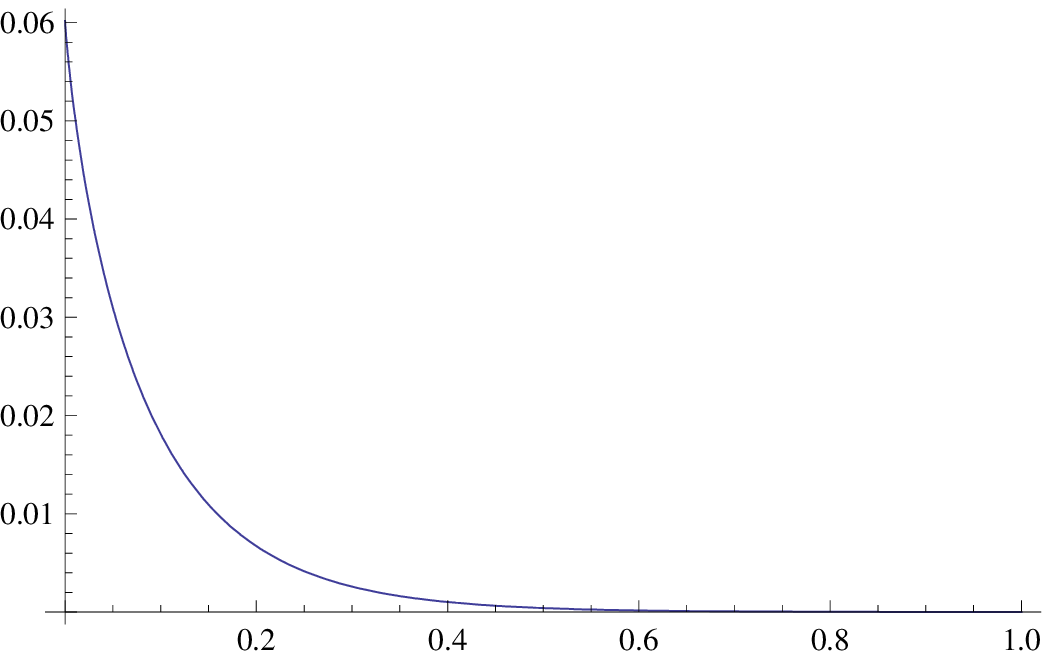}
\includegraphics[width=6cm]{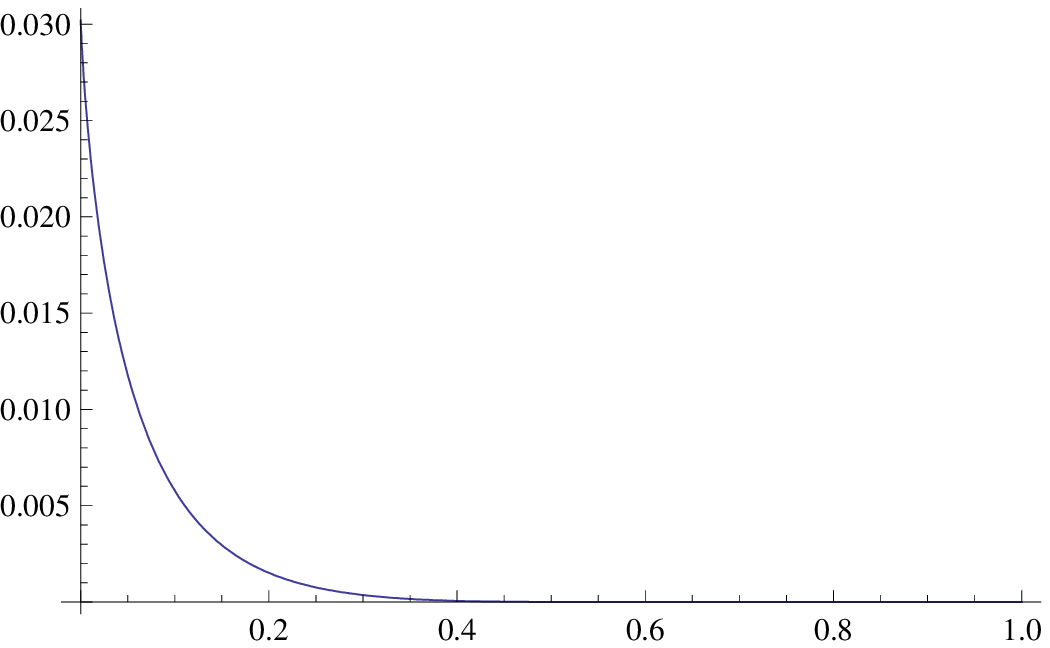}
\caption{Case 4: $\alpha=0.5$; Left: $S(\varrho_t||\varrho_\infty)$; Right:
$E[\varrho_t]$} \label{fig24}
\end{figure}

\begin{figure}[h!]
\centering
\includegraphics[width=6cm]{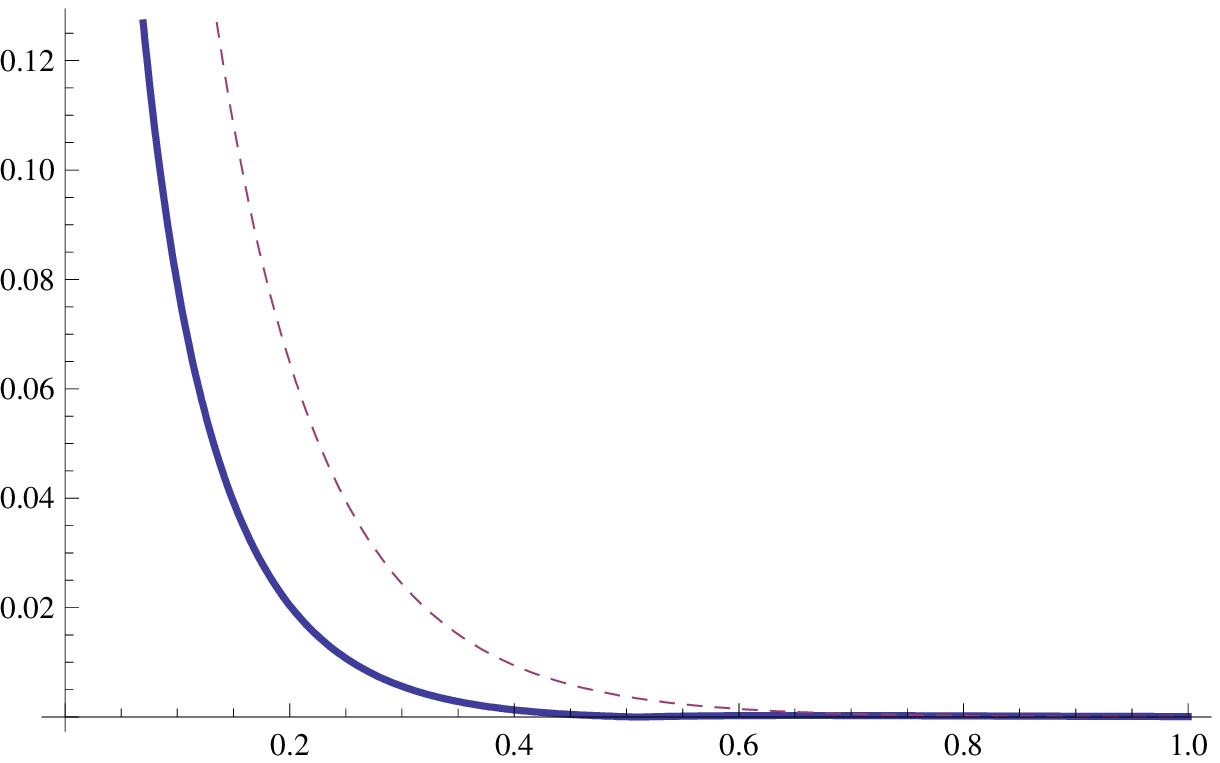}
\caption{Case 4: $\alpha=0.5$; $\sigma[\varrho_t]$ dashed line,
$|\sigma_E[\varrho_t]|$ continuous line} \label{fig6}
\end{figure}

\newpage
\noindent \textbf{Case 5.}\quad An initial mixed entangled
state~(\ref{state4}) goes into a mixed entangled state with more or
less entanglement depending on the choice of the parameter $\alpha$.
\begin{eqnarray}
\label{state4}
&&
\varrho=
\frac{3}{10}\vert 2\rangle\langle 2\vert\,+\,\frac{1}{10}\,\vert 3\rangle\langle 3\vert\,
+\,\frac{3}{5}\,\vert 4\rangle\langle 4\vert\\
\nonumber
&&
\varrho_\infty=
\frac{2(1-\alpha)^2}{5(3+\alpha^2)}\,\vert 1\rangle\langle 1\vert\,+\,
\frac{2(1+\alpha)^2}{5(3+\alpha^2)}\,\vert 2\rangle\langle 2\vert\,+\,
\frac{2(1-\alpha^2)}{5(3+\alpha^2)}\,\vert 3\rangle\langle 3\vert\,+\,\frac{3}{5}\,\vert 4\rangle\langle 4\vert\\
\nonumber &&C(\varrho)=\frac{1}{2}\ ,\quad
C(\varrho_\infty)=\frac{3(1+3\alpha^2)}{5(3+\alpha^2)}
\end{eqnarray}

The concurrence of the asymptotic state $C(\varrho_\infty)$ can be
larger or smaller than $C(\varrho)=\frac{1}{2}$ depending on the
value of the parameter $\alpha$.

If, for instance, we take $\alpha=0.5$, then
$C(\varrho_\infty)<\frac{1}{2}$, i.e. the asymptotic state has less
entanglement than the initial state, and from the plot of the
entanglement rate vs. the entropy rate (Figure 11) we can see that
the conjecture~(\ref{vedralconj}) is always violated.
\begin{figure}[h!]
\centering
\includegraphics[width=6cm]{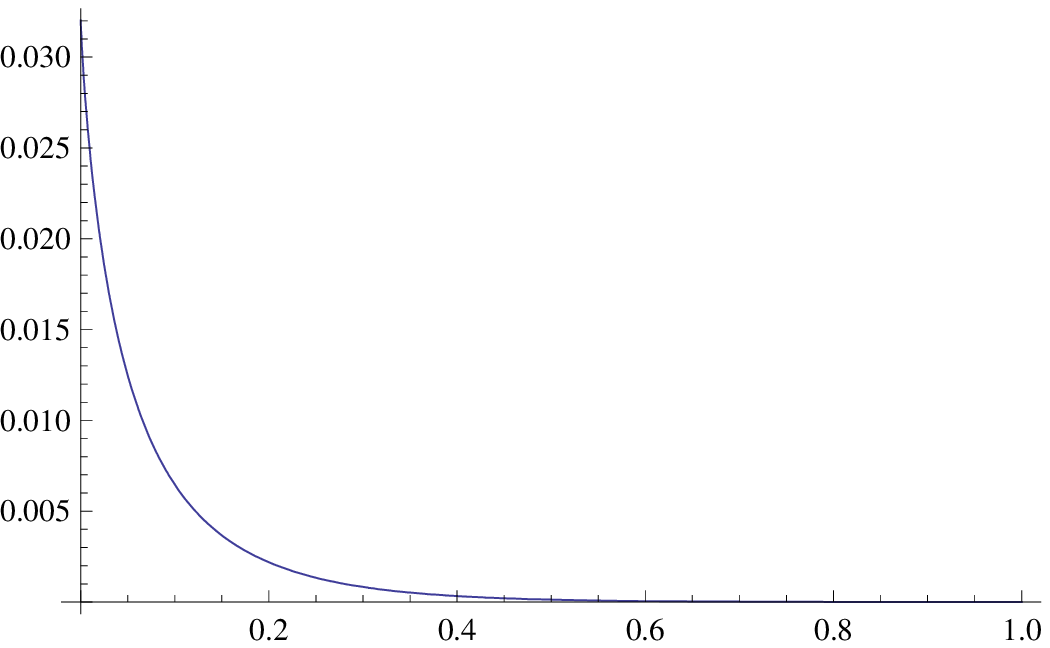}
\includegraphics[width=6cm]{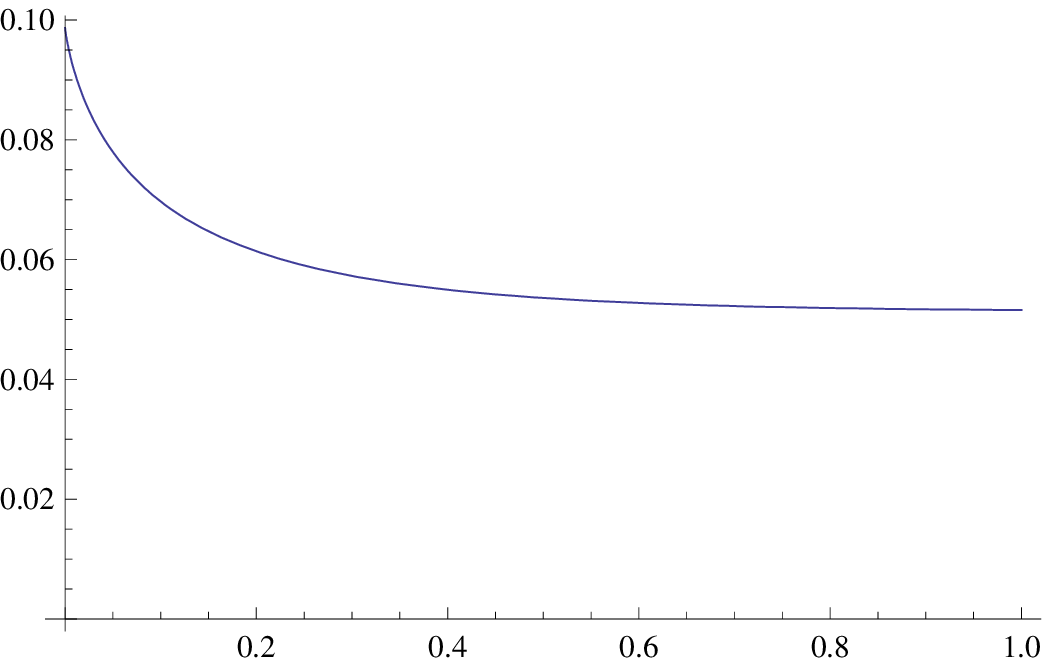}
\caption{Case 5: $\alpha=0.5$; Left: $S(\varrho_t||\varrho_t)$; Right:
$E[\varrho_t]$} \label{fig235}
\end{figure}

\begin{figure}[h!]
\centering
\includegraphics[width=6cm]{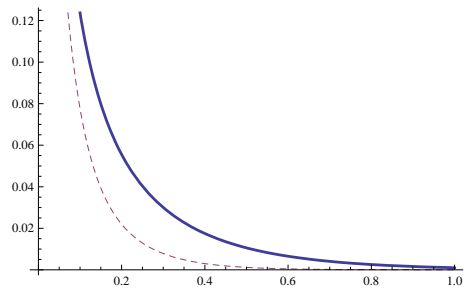}
\caption{Case 5: $\alpha=0.5$; $\sigma[\varrho_t]$ dashed line,
$|\sigma_E[\varrho_t]|$ continuous line} \label{fig4}
\end{figure}

If, instead, we take for example $\alpha=0.8$, then the initial
entanglement first diminishes and then increases again, leading to
an asymptotic state with more entanglement than the initial one, as
can be seen from the plot of the entropy of entanglement as a
function of time (Figure 12). From the corresponding plot of the
entanglement rate vs. the entropy rate (Figure 13) we can see that
in this case the conjecture~(\ref{vedralconj}) is violated after
some time.
\begin{figure}[h!]
\centering
\includegraphics[width=6cm]{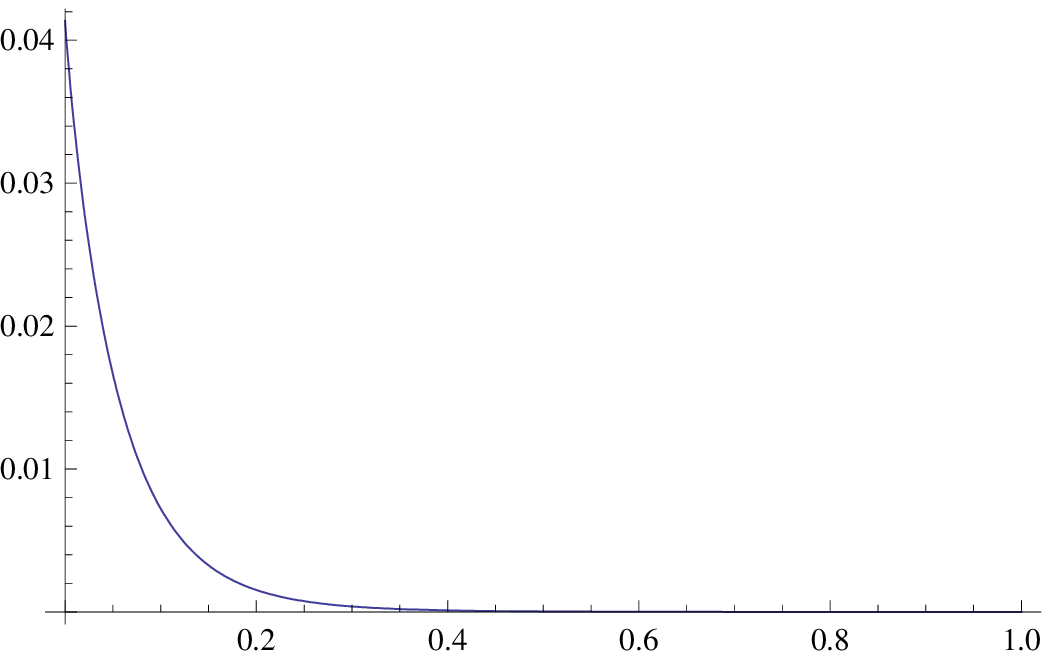}
\includegraphics[width=6cm]{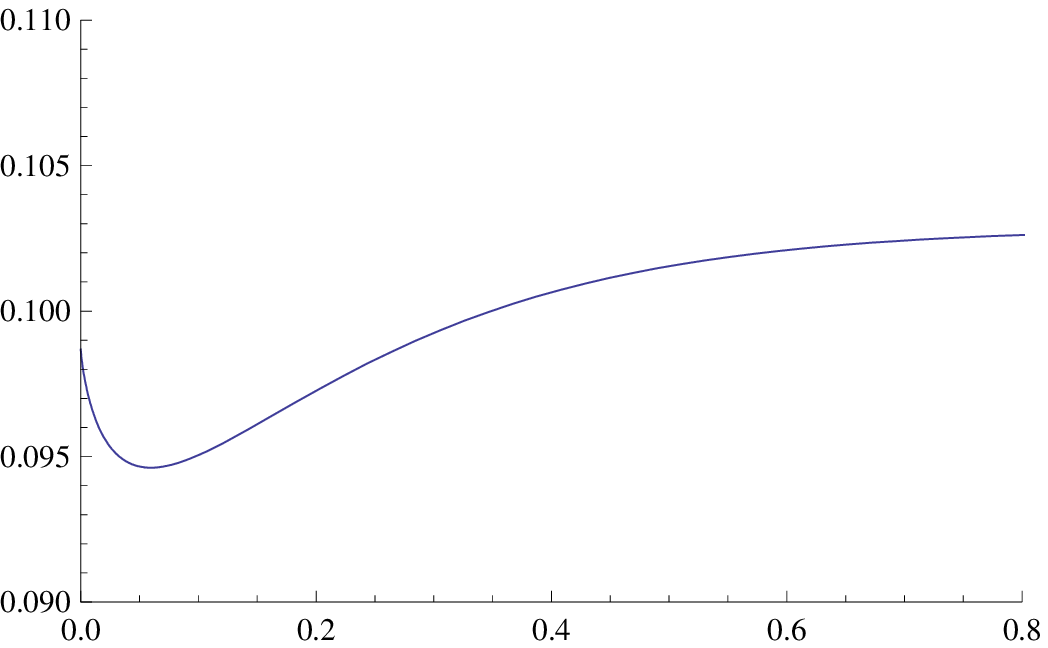}
\caption{Case 5: $\alpha=0.8$; Left: $S(\varrho_t||\varrho_\infty)$; Right:
$E[\varrho_t]$} \label{fig238}
\end{figure}

\begin{figure}[h!]
\centering
\includegraphics[width=6cm]{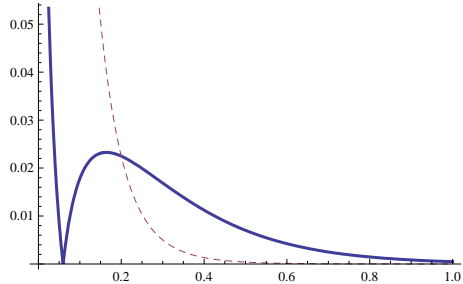}
\caption{Case 5: $\alpha=0.8$; $\sigma[\varrho_t]$ dashed line,
$|\sigma_E[\varrho_t]|$ continuous line} \label{fig11}
\end{figure}

%\newpage
\noindent {\bf Remark 4}\quad In the last plot of the entanglement
rate vs. the entropy rate, the cusp is due to the change of sign in
the entropy of entanglement $S(\varrho_t||\varrho_\infty)$ and to
the fact that in the conjecture~(\ref{vedralconj}) the absolute
value of the entanglement rate $\sigma_E[\varrho_t]$ is considered.
On the other hand, all the other plots of the entanglement rate
present a continuous behavior which reflects the fact that the
entropy of entanglement does not change sign, i.e. it either
increases or decreases monotonically. Finally, the plots of the
relative entropy show its monotonic behavior under the action of
completely positive trace-preserving maps; and to this corresponds a
monotonic decreasing behavior for the entropy rate.

\section{Conclusions}

The time-derivative of the quantum relative entropy serves as a
measure of how fast an open quantum system tends to equilibrium
dissipating free energy under a quantum dynamical semigroup of
completely positive maps generated by a Lindblad-type master
equation. On the other hand, via a variational formulation, the
relative entropy may be used as a pseudo-distance of an entangled
bipartite state from the convex subset of separable states (relative
entropy of entanglement); therefore, its time derivative can be
interpreted as the speed with which a time-evolving state moves
toward, or away from, becoming separable.

Based on the expectation that the entanglement content of
dissipatively-driven bipartite systems disappears asymptotically due
to decoherence effects, in~\cite{vedral} a conjecture was put
forward, namely that the entropy rate, measured by (minus) the
time-derivative of the relative entropy of a dissipatively
evolving state and its asymptotic state, should
always be larger than the absolute value of the time-derivative of
the relative entropy of entanglement.

However, beside being a source of decoherence, an environment can in
some cases build quantum correlations that can even persist
asymptotically: in this paper we have studied the fate of the above
conjecture in the case of a Lindblad-type master equation that
presents a rich manifold of asymptotic states that may be more or
less entangled with respect to the initial states they emerge from.
The entropy and entanglement rates have been explicitly calculated
and numerically plotted for a class of initial states. It turns out
that, when the asymptotic state is entangled, the conjecture is
violated either at all times or after a finite time; instead, the
conjecture is confirmed in all cases when the asymptotic state is
separable. The conjecture put forward in~\cite{vedral} should be
thus reformulated as follows:
\begin{equation}
\label{vedralconjref}
\Big|\sigma_E[\varrho_t]\Big| \leq \sigma[\rho_t]\quad\forall \rho_t\ \hbox{with}\ \varrho_\infty\
\hbox{separable}\ .
\end{equation}
While the asymptotic predominance of the entropy rate over the
entanglement rate in the latter case has already been explained
in~\cite{vedral} based on the fact that there are no entangled
states in a suitable neighborhood of the separable asymptotic state,
the truly remarkable fact about~(\ref{vedralconjref}) is its
validity at all $t\geq 0$ in all the cases that have been checked.

\section{Appendix A}

In order to explicitly solve the master equation~(\ref{diag-noise}),
one first writes the matrix $A$ in diagonal form:
$$
A=U\,\begin{pmatrix}
1+\alpha&0&0\cr0&1-\alpha&0\cr0&0&1
\end{pmatrix}\ U^\dagger\ ,\qquad U =
\begin{pmatrix}

                 1/\sqrt{2}  & 1/\sqrt{2} & 0 \\

                 -i/\sqrt{2} & i/\sqrt{2} & 0 \\

                 0  & 0 & 1
\end{pmatrix}\ ,
$$
and recasts the dissipative term in~(\ref{diss-gen}) in the form
\begin{eqnarray}
\label{diag-noise}
\mathbf{D}[\varrho_t]&=&2(1+\alpha)\Big(\Sigma_- \varrho_t \Sigma_+
                     - \frac{1}{2}\{\Sigma_+\Sigma_-, \varrho_t\}\,+\,
                     2(1-\alpha)\Big(\Sigma_+ \varrho_t \Sigma_-
                     - \frac{1}{2}\{\Sigma_-\Sigma_+, \varrho_t\}\Big) \nonumber \\
                     & + & \Sigma_3 \varrho_t \Sigma_3 - \frac{1}{2}\{\Sigma_3\Sigma_3, \varrho_t\}\ ,
\end{eqnarray}
with $\Sigma_{\pm} := \frac{1}{2}(\Sigma_1 \pm i\Sigma_2)$. Since
$\displaystyle H=\frac{\Omega}{2}\,\Sigma_3$, it follows that
$\displaystyle {\rm e}^{iHt}\Sigma_\pm{\rm e}^{-iHt}={\rm e}^{\pm
i\Omega t}\,\Sigma_\pm$. Thus, setting $\widetilde{\varrho}_t :=
e^{iHt}\varrho_t e^{-iHt}$, in the interaction picture the master
equation~(\ref{diag-noise}) becomes
\begin{eqnarray}
\label{tilde-master-eq}
\partial_t\widetilde{\varrho}_t & = & {\rm e}^{iHt}\mathbf{D}[{\rm e}^{-iHt}\tilde{\varrho}_t{\rm e}^{iHt}]e^{-iHt} \nonumber \\
& = & 2(1+\alpha)\Big(\Sigma_- \tilde{\varrho}_t \Sigma_+
                      - \frac{1}{2}\{\Sigma_+\Sigma_-, \tilde{\varrho}_t\}\Big)\,+\,
                      2(1-\alpha)\Big(\Sigma_+ \tilde{\varrho}_t\Sigma_-
                      - \frac{1}{2}\{\Sigma_-\Sigma_+, \tilde{\varrho}_t\}\Big) \nonumber \\
                     & + & \Sigma_3 \tilde{\varrho}_t\Sigma_3 - \frac{1}{2}\{\Sigma_3\Sigma_3, \tilde{\varrho}_t\}\ .
\end{eqnarray}
In order to solve it, it proves convenient to represent
$\widetilde{\rho}_t=\sum_{i,j=1}^4\rho_{ij}(t)\vert i\rangle\langle
j\vert$ with respect to the orthonormal basis~(\ref{ONB}). Indeed,
using that
\begin{equation}
\label{A1}
\left\{
\begin{matrix}
\Sigma_+\vert 1\rangle&=&0\cr
\Sigma_+\vert 2\rangle&=&\sqrt{2}\vert 3\rangle\cr
\Sigma_+\vert 3\rangle&=&\sqrt{2}\vert 1\rangle\cr
\Sigma_+\vert 4\rangle&=&0\cr
\end{matrix}\right.\ ,\quad
\left\{
\begin{matrix}
\Sigma_-\vert 1\rangle&=&\sqrt{2}\vert 3\rangle\cr
\Sigma_-\vert 2\rangle&=&0\cr
\Sigma_-\vert 3\rangle&=&\sqrt{2}\vert 2\rangle\cr
\Sigma_-\vert 4\rangle&=&0\cr
\end{matrix}\right.\ ,\quad
\left\{
\begin{matrix}
\Sigma_3\vert 1\rangle&=&2\vert 1\rangle\cr
\Sigma_3\vert 2\rangle&=&-2\vert 2\rangle\cr
\Sigma_3\vert 3\rangle&=&0\cr
\Sigma_3\vert 4\rangle&=&0\cr
\end{matrix}\right.\ ,
\end{equation}
one derives from~(\ref{tilde-master-eq}) the following equations :
$$
\begin{matrix}
\dot{\rho}_{11}&=&-4(1+\alpha)\rho_{11}+4(1-\alpha)\rho_{33}&,&
\dot{\rho}_{12}&=&-12\rho_{12}\cr
\dot{\rho}_{13}&=&-2(4+\alpha)\rho_{13}+4(1-\alpha)\rho_{32}&,&
\dot{\rho}_{14}&=&-2(2+\alpha)\rho_{14}\cr
\dot{\rho}_{22}&=&-4(1-\alpha)\rho_{22}+4(1+\alpha)\rho_{33}&,&
\dot{\rho}_{23}&=&-2(4-\alpha)\rho_{23}+4(1+\alpha)\rho_{31}\cr
\dot{\rho}_{33}&=&4(1+\alpha)\rho_{11}+4(1-\alpha)\rho_{22}-8\rho_{33}&,&
\dot{\rho}_{24}&=&-2(2-\alpha)\rho_{24}\cr
\dot{\rho}_{34}&=&-4\rho_{34}&,&\dot{\rho}_{44}&=&0\ ,
\end{matrix}
$$
plus the complex conjugated equations for $\dot{\rho}_{ij}$, $i\neq j$; whence
\begin{equation}
\label{r00}
\begin{matrix}
\rho_{12}(t)&=&\rho_{12}\,{\rm e}^{-12 t}&,&
\rho_{14}(t)&=&\rho_{14}\,{\rm e}^{-2(2+\alpha)t}\cr
\rho_{24}(t)&=&\rho_{24}\,{\rm e}^{-2(2-\alpha)t}&,&
\rho_{34}(t)&=&\rho_{34}\,{\rm e}^{-4t}\cr
\rho_{44}(t)&=&\rho_{44}\ .&&&&
\end{matrix}
\end{equation}
Of the remaining equations, two of them couple the off-diagonal
terms $\rho_{13}$ and $\rho_{32}$, yielding
\begin{eqnarray}
\label{r13}
\rho_{13}(t)&=&\rho_{13}\;F_+(t)\,+\,\frac{2(1-\alpha)\rho_{32}-\alpha\rho_{13}}{\sqrt{4-3\alpha^2}}\;
F_-(t)\\
\label{r32}
\rho_{32}(t)&=&\rho_{32}\;F_+(t)\,+\,\frac{2(1+\alpha)\rho_{13}+\alpha\rho_{32}}{\sqrt{4-3\alpha^2}}\;
F_-(t)\ ,
\end{eqnarray}
while the other three solutions couple the diagonal entries:
\begin{eqnarray}
\nonumber
\rho_{11}(t)&=&\frac{(1-\alpha)^2}{3+\alpha^2}\;R\,+\,\sqrt{1-\alpha^2}\,
\frac{(1+\alpha)^2\rho_{11}-2(1-\alpha)\rho_{22}\,+\,
(1+\alpha)^2\rho_{33}}{(1+\alpha)(3+\alpha^2)}\;E_-(t)\\
&+&
\frac{2(1+\alpha)\rho_{11}-(1-\alpha)^2(\rho_{22}+\rho_{33})}{3+\alpha^2}\;E_+(t)
\label{r11}
\\
\nonumber
\rho_{22}(t)&=&\frac{(1+\alpha)^2}{3+\alpha^2}\;R\,-\,\sqrt{1-\alpha^2}\,\frac{2(1+\alpha)\rho_{11}
-(1-\alpha)^2(\rho_{22}+\rho_{33})}{(1-\alpha)(3+\alpha^2)}\;\ E_-(t)\\
&-&
\frac{(1+\alpha)^2\rho_{11}-2(1+\alpha)\rho_{22}+(1+\alpha)^2\rho_{33}}{3+\alpha^2}\;E_+(t)
\label{r22}\\
\nonumber
\rho_{33}(t)&=&\frac{(1-\alpha^2)}{3+\alpha^2}\;R\,+\,\sqrt{1-\alpha^2}\,\frac{(1+\alpha)^3\rho_{11}
+(1-\alpha)^3\rho_{22}-2(1-\alpha^2)\rho_{33}}{(3+\alpha^2)(1-\alpha^2)}\;E_-(t)\\
\label{r33}
&+&
\frac{2(1+\alpha^2)\rho_{33}-(1-\alpha^2)(\rho_{11}+\rho_{22})}{3+\alpha^2}\;E_+(t)\ ,
\end{eqnarray}
where
$R=\rho_{11}+\rho_{22}+\rho_{33}=\rho_{11}(t)+\rho_{22}(t)+\rho_{33}(t)$
is a constant of the motion and
\begin{equation}
E_\pm(t)={\rm e}^{-8t}\,\left\{
\begin{matrix}
\cosh{4t\sqrt{1-\alpha^2}}\cr
\sinh{4t\sqrt{1-\alpha^2}}
\end{matrix}\right.\ ,\quad
F_\pm(t)={\rm e}^{-8t}\,\left\{
\begin{matrix}
\cosh{2t\sqrt{4-3\alpha^2}}\cr
\sinh{2t\sqrt{4-3\alpha^2}}
\end{matrix}\right.\ .
\label{r44}
\end{equation}
are quantities which decay asymptotically with $t\to+\infty$.
The remaining entries $\rho_{ij}(t)$ follow from complex conjugation.
By returning to the Schr\"odinger representation, using~(\ref{A1})
the explicit solution of~(\ref{diag-noise}) reads
\begin{eqnarray}
\label{A2}
\varrho_t=\sum_{i,j=1}^4\varrho_{ij}(t)\
{\rm e}^{2i\omega\,t(\delta_{j1}+\delta_{i2}-\delta_{i1}-\delta_{j2})}\
\vert i\rangle\langle j\vert\ .
\end{eqnarray}

\section{Appendix B}

In order to prove the Proposition in Section 3, let us consider the
spectral decompositions $\varrho_t = \sum_{i=1}^4
r_i(t)|i\rangle\langle i|$ (see~(\ref{rho-tilde})) and
$\varrho_{sep}=\sum_{j=1}^4 s_j|s_j\rangle\langle s_j|$. We have:
\begin{eqnarray}
\label{log}
{\rm Tr}\Big(\varrho_t\log\varrho_{sep}\Big) & = & \sum_{i=1}^4
r_i(t)\langle i|\log\varrho_{sep}|i\rangle
                                 = \sum_{i=1}^4 r_i(t)\sum_{j=1}^4|\langle
                                 i|s_j\rangle|^2\,\log s_j \nonumber \\
                                 & \leq & \sum_{i=1}^4 r_i(t)\log\Big(\sum_{j=1}^4 s_j|\langle
                                 i|s_j\rangle|^2\Big) \nonumber \\
                                 & = &\sum_{i=1}^4 r_i(t)\log\langle
                                 i|\varrho_{sep}|i\rangle =
                                 {\rm Tr}\Big(\varrho_t\log\Pi[\varrho_{sep}]\Big)\ ,
\end{eqnarray}
where the inequality follows from the convexity of $\log x$,
$$
\log \sum_i\lambda_i\,x_i\geq\sum_i\lambda_i\,\log x_i\ ,\qquad
\lambda_i\geq 0\, ,\quad \sum_i\lambda_i=1\ ,\quad x_i\geq 0\ ,
$$
and $\sum_{j=1}^4|\langle i\vert s_j\rangle|^2=1$.
Also, we have introduced the completely positive map
\begin{equation}
\label{CPrho}
\varrho\mapsto\Pi[\varrho]:=\sum_{i=1}^4|i\rangle\langle
i|\varrho|i\rangle\langle i|\ ,
\end{equation}
on the 2-qubit density matrices $\mathcal{S}(\mathbb{C}^4)$ that
diagonalizes its argument with respect to the orthonormal
basis~(\ref{ONB}). This map has the following property which allows
one to analytically solve the variational problem~(\ref{log}).
\medskip

\noindent
\textbf{Lemma}\quad
$\Pi:\mathcal{S}(\mathbb{C}^4)\mapsto\mathcal{S}(\mathbb{C}^4)$ maps separable states into separable states.
\medskip

\noindent
\textbf{Proof:}\quad
Given the density matrix of an arbitrary
$2$-qubit state in the standard basis
$$
\varrho=
\begin{pmatrix}

                 \varrho_{00,00}  & \varrho_{00,01} & \varrho_{00,10} & \varrho_{00,11}  \\

                 \varrho_{01,00} & \varrho_{01,01} & \varrho_{01,10} & \varrho_{01,11} \\

                 \varrho_{10,00}  & \varrho_{10,01} & \varrho_{10,10} & \varrho_{10,11} \\

                 \varrho_{11,00} & \varrho_{11,01} & \varrho_{11,10} & \varrho_{11,11}
\end{pmatrix}\ ,
$$
the action of the map $\Pi$ transforms it into a density matrix of the form
$$
\Pi[\varrho]=\begin{pmatrix}

                 \varrho_{00,00}  & 0 & 0 & 0  \\

                 0 & \frac{\varrho_{01,01}+\varrho_{10,10}}{2} & \Re{\rm e}\big(\varrho_{01,10}\big) & 0 \\

                 0  & \Re{\rm e}\big(\varrho_{01,10}\big) & \frac{\varrho_{01,01}+\varrho_{10,10}}{2} & 0 \\

                 0 & 0 & 0 & \varrho_{11,11}
\end{pmatrix}\ .
$$
By partial transpostion~\cite{horodecki}, $\Pi[\varrho]$ is
entangled if and only if $|\Re{\rm e}(\varrho_{01,10})|\geq
\sqrt{\varrho_{00,00}\varrho_{11,11}}$. But then, the partially
transposed $\varrho$ (with respect to the second qubit),
$$
\varrho^\Gamma=
\begin{pmatrix}

                 \varrho_{00,00}  & \varrho_{01,00} & \varrho_{00,10} & \varrho_{01,10}  \\

                 \varrho_{00,01} & \varrho_{01,01} & \varrho_{00,11} & \varrho_{01,11} \\

                 \varrho_{10,00}  & \varrho_{11,00} & \varrho_{10,10} & \varrho_{11,10} \\

                 \varrho_{10,01} & \varrho_{11,01} & \varrho_{10,11} & \varrho_{11,11}
\end{pmatrix}
$$
cannot be positive semi-definite, for $|\varrho_{01,10}|\geq|\Re{\rm e}(\varrho_{01,10})|>\sqrt{\varrho_{00,00}\varrho_{11,11}}$ in the sub-matrix
$\displaystyle
\begin{pmatrix}
\varrho_{11}&\varrho_{01,10}\cr
\varrho_{10,01}&\varrho_{22}
\end{pmatrix}$.
Therefore, if $\varrho$ is separable, then also $\Pi[\varrho]$ must
be so. \hfill$\Box$
\bigskip

Observe that~(\ref{log}) implies
$\sup_{\varrho_{sep}}{\rm Tr}\Big(\varrho_t\log\varrho_{sep}\Big)
\leq
\sup_{\varrho_{sep}}{\rm Tr}\Big(\varrho_t\log\Pi[\varrho_{sep}]\Big)$; on the other hand,
since $\Pi$ maps separable states into separable states,
$$
\sup_{\varrho_{sep}}{\rm Tr}\Big(\varrho_t\log\varrho_{sep}\Big)
\leq\sup_{\varrho_{sep}}{\rm Tr}\Big(\varrho_t\log\Pi[\varrho_{sep}]\Big)
\leq \sup_{\varrho_{sep}}{\rm Tr}\Big(\varrho_t\log\varrho_{sep}\Big)\ .
$$
Thus, the maximum in~(\ref{log}) is attained on the subset
$\mathcal{S}_{sep}^{diag}$ of separable qubit states that are
diagonal with respect to the orthonormal basis, namely of the
form~(\ref{constr1}) with the second bound on the real parameters
$x,y,u,v$ in~(\ref{param}) coming from the condition of positivity
under partial transposition of matrices of the form~(\ref{matform}),
which is necessary and sufficient for separability.

It thus follows from~(\ref{constr0}) and ~(\ref{log}) that for
$\varrho_t$ as in~(\ref{rho-tilde}) the relative entropy of
entanglement can be reduced to the computation of~(\ref{max}). In
order to explicitly solve such a variational problem, we seek the
stationary points of a function of the form
$$
f(x,y,u,v) := a\log x + d \log y +
b \log u + c \log v + \lambda(x+y+u+v-1)
$$
with given $a,b,c,d\geq 0$ such that $a+b+c+d=1$, relative to variations of the parameters
$x,y,u,v$ over values achieving separable states
of the form~(\ref{constr1}).
Stationarity implies
$$
a = -\lambda x, \quad d = -\lambda y, \quad b = -\lambda u, \quad c
= -\lambda v\ ;
$$
whence $\lambda=-1$ and $a=x, d=y, b=u, c=v$. However, this can be
the required solution only if the state $\varrho$ in $E[\varrho]$ is
separable so that $E[\varrho]=0$. Otherwise, the solution must lie
on the border of the subset of separable states of the
form~(\ref{constr1}), where the inequality in~(\ref{param}) is
saturated. From $x+u+v+y=1$ and $(u-v)^2=4\,x\,y$, one gets
$$
u_\pm = \frac{1 - (\sqrt{x} \mp \sqrt{y})^2}{2} , \qquad v_\pm =
\frac{1 - (\sqrt{x} \pm \sqrt{y})^2}{2}\ ,
$$
so that the function to be maximized becomes
\begin{equation}
\label{funct} f(x,y) = a \log x + d \log y + b \log \big(\frac{1 -
(\sqrt{x} \mp \sqrt{y})^2}{2}\big) + c \log \big(\frac{1 - (\sqrt{x}
\pm \sqrt{y})^2}{2}\big) .
\end{equation}
Stationarity with respect to $x,y$ leads to a system of two equations for
the two unknowns $x$ and $y$ in terms of the coefficients $a, b, c, d$.
From setting $\partial_{x,y}f(x,y)=0$ and from the condition $a+b+c+d=1$ it follows that
$y=x-a+d$ and that
\begin{eqnarray*}
x & = & \frac{1}{8 (-1 + b) (a + b + d)}\Big\{ -a^3 -d(-1+2b+d)^2
+ a^2(-6+4b+d) \\
& + & a(-1 +4(-1+b) +d^2)
+ \Big[(-1+a+2b+d)^2\Big(a^4 + 2a^3(-1+2b-2d) \\
& + & d^2(-1+2b+d)^2 + a^2(1+4b^2+2d+6d^2-4b(1+d)) \\ & + &
2ad(1+d-2(2b(-1+b) + bd + d^2))\Big)\Big]^{1/2}\Big\}.
\end{eqnarray*}
By inserting into it the values~(\ref{sa})--(\ref{sc}), this
expression yields the separable state of the form~(\ref{constr1})
which is closest to an evolving entangled state of the
form~(\ref{rho-tilde}). Though cumbersome, the resulting
entanglement rate~(\ref{ent-prod}) is amenable to numerical
inspection.


\begin{thebibliography}{99}

\bibitem{Bruss}
D. Bruss, G. Leuchs, {\em Lectures on quantum information}, (Wiley-Vch 2007)

\bibitem{AL}
R. Alicki and K. Lendi, {\em Quantum Dynamical Semigroups and
Applications}, Lect. Notes Phys. {\bf 286}, (Springer-Verlag, Berlin, 1987)

\bibitem{spohn}
H. Spohn, Rev. Mod. Phys. {\bf 52}, 569 (1980)

\bibitem{BP}
H.-P. Breuer, F. Petruccione, {\em The Theory of Open
Quantum Systems} (Oxford University Press, Oxford, 2002)

\bibitem{GFK}
V. Gorini, A. Frigerio, M. Verri et al., Rep. Math. Phys. {\bf 13}, 149 (1978)

\bibitem{Lindblad}
G. Lindblad, Commun. Math. Phys. {\bf 48}, 119 (1976)


\bibitem{ben-flore} F. Benatti, R. Floreanini,
Int. J. Mod. Phys. B \textbf{19}, 19 (2005)


\bibitem{OP}
M. Ohya, D. Petz, \textit{Quantum Entropy and Its Use}, Springer, Berlin 1993

\bibitem{plenio-vedral} M. B. Plenio, V. Vedral, Cont. Phys. \textbf{39}, 431 (1998)

\bibitem{vedral} V. Vedral, Journal of Physics: Conference Series \textbf{143} (2009) 012010

\bibitem{beige} A. Beige \textit{et al.}, J. Mod. Opt. \textbf{47},
2583 (2000)

%\bibitem{plenio-huelga} M. B. Plenio, S. F. Huelga, Phys. Rev. Lett.
%\textbf{88}, 197901 (2002)

\bibitem{braun} D. Braun, Phys. Rev. Lett.
\textbf{89}, 277901 (2002)

\bibitem{jacob} L. Jacobczyk, J. Phys. A: Math. gen. \textbf{35},
6383 (2002)

\bibitem{palma} M. A. Cirone, G. M. Palma, Advanced Science Letters, Vol. 2, 1-3
(2009)


\bibitem{BFP}
F. Benatti, R. Floreanini, M. Piani, Phys. Rev. Lett. \textbf{91}, 070402 (2003)

\bibitem{liguori-nagy} F. Benatti, Alexandra M. Liguori, A. Nagy, J. Math. Phys.
\textbf{49}, 042103 (2008)

\bibitem{vprk} V. Vedral, M. B. Plenio, M. A. Rippin, P. L. Knight,
Phys. Rev. Lett. \textbf{78}, 2275 (1997)

\bibitem{vedral-plenio} V. Vedral, M. B. Plenio,
Phys. Rev. A \textbf{57}, 1619 (1998)

\bibitem{Wootters}
W.K. Wootters, Phys. Rev. Lett. \textbf{80}, 2245 (1998)

\bibitem{horodecki} M. Horodecki, P. Horodecki, R. Horodecki,
Phys. Lett. A \textbf{223}, 1 (1996)



\end{thebibliography}
\end{document}